\documentclass[trackchanges]{aastex7}

\begin{document}

\title{Orbital and Pulsation Analysis of 42 Heartbeat Stars Discovered in TESS Data}

\author[orcid=0000-0002-8564-8193]{Min-Yu Li}
\affiliation{Yunnan Observatories, Chinese Academy of Sciences, Kunming 650216, People's Republic of China}
\email{liminyu@ynao.ac.cn} 

\author[orcid=0000-0002-5995-0794]{Sheng-Bang Qian}
\affiliation{Department of Astronomy, School of Physics and Astronomy, Key Laboratory of Astroparticle Physics of Yunnan Province, Yunnan University, Kunming 650091, People's Republic of China}
\email[show]{qiansb@ynu.edu.cn}

\author[orcid=0000-0002-2919-6989]{Ai-Ying Zhou}
\affiliation{National Astronomical Observatories, Chinese Academy of Sciences, A20 Datun Road, Chaoyang District, Beijing 100101, People's Republic of China}
\email{aiying@nao.cas.cn}

\author[orcid=0000-0002-0796-7009]{Li-Ying Zhu} 
\affiliation{Yunnan Observatories, Chinese Academy of Sciences, Kunming 650216, People's Republic of China}
\affiliation{University of Chinese Academy of Sciences, No.1 Yanqihu East Road, Huairou District, Beijing 101408, People's Republic of China}
\email[show]{zhuly@ynao.ac.cn}

\author[0000-0001-9346-9876]{Wen-Ping Liao} 
\affiliation{Yunnan Observatories, Chinese Academy of Sciences, Kunming 650216, People's Republic of China}
\affiliation{University of Chinese Academy of Sciences, No.1 Yanqihu East Road, Huairou District, Beijing 101408, People's Republic of China}
\email{liaowp@ynao.ac.cn}

\author[orcid=0000-0002-8421-4561]{Lin-Feng Chang}
\affiliation{Department of Astronomy, School of Physics and Astronomy, Key Laboratory of Astroparticle Physics of Yunnan Province, Yunnan University, Kunming 650091, People's Republic of China}
\email{chang-linfeng@ynu.edu.cn}

\author{Er-Gang Zhao} 
\affiliation{Yunnan Observatories, Chinese Academy of Sciences, Kunming 650216, People's Republic of China}
\email{zergang@ynao.ac.cn}

\author[orcid=0000-0002-5038-5952]{Xiang-Dong Shi} 
\affiliation{Yunnan Observatories, Chinese Academy of Sciences, Kunming 650216, People's Republic of China}
\email{sxd@ynao.ac.cn}

\author[orcid=0000-0002-0285-6051]{Fu-Xing Li}
\affiliation{Department of Astronomy, School of Physics and Astronomy, Key Laboratory of Astroparticle Physics of Yunnan Province, Yunnan University, Kunming 650091, People's Republic of China}
\email{lfxjs66@126.com}

\author[orcid=0000-0003-0516-404X]{Qi-Bin Sun}
\affiliation{Department of Astronomy, School of Physics and Astronomy, Key Laboratory of Astroparticle Physics of Yunnan Province, Yunnan University, Kunming 650091, People's Republic of China}
\email{sunqibin@ynu.edu.cn}

\author[0009-0004-0289-2732]{Ping Li} 
\affiliation{Yunnan Observatories, Chinese Academy of Sciences, Kunming 650216, People's Republic of China}
\affiliation{University of Chinese Academy of Sciences, No.1 Yanqihu East Road, Huairou District, Beijing 101408, People's Republic of China}
\email{liping@ynao.ac.cn}


\begin{abstract}
Heartbeat stars (HBSs) are ideal laboratories for studying the formation and evolution of binary stars in eccentric orbits and their mutual tidal interactions. We present 42 new HBSs discovered based on TESS-SPOC and QLP data. Their physical parameters have been obtained through modeling with appropriate models. Subsequently, Tidally excited oscillations (TEOs) are detected in ten systems, and their pulsation phases and modes are identified. Most pulsation phases can be explained by the dominant being spherical harmonic degree $l=2$ and azimuthal order $m=0$ or $\pm2$. For TIC 156846634, the harmonic with large deviation ($>3\sigma$) from the expected adiabatic phase can be expected to be a traveling wave or significantly nonadiabatic. The harmonic numbers $n$ = 16 in TIC 184413651 may not be considered as a TEO candidate due to its large deviation ($>2\sigma$) from the adiabatic expectation. Moreover, TIC 92828790 shows no TEOs but exhibits a significant $\gamma$\,Dor-type pulsation. The eccentricity-period ($e-P$) relation also shows a positive correlation between eccentricity and period, as well as the existence of orbital circularization. The Hertzsprung-Russell diagram shows that TESS HBSs have higher temperatures and greater luminosities than Kepler HBSs, possibly due to selection effects. This significantly enhances the detectability of massive HBSs and those containing TEOs.

\end{abstract}



\section{Introduction}\label{sect:introduction}

Heartbeat stars (HBSs) are a subclass of ellipsoidal variables. They are detached binaries with eccentric orbits. The term originates from the characteristic ``heartbeat''-like features observed in their light curves, which are analogous to electrocardiogram patterns\citep{2012ApJ...753...86T}. The components are distorted by the time-varying tidal potential, and their response is usually divided into two parts: the equilibrium tide and the dynamical tide. The equilibrium tide and effects such as reflection, irradiation, and Doppler boosting are responsible for the heartbeat signature near periastron, while the dynamical tide is composed of tidally excited oscillations (TEOs; \citet{1975A&A....41..329Z,1995ApJ...449..294K,2017MNRAS.472.1538F}). HBSs are intriguing objects that are ideal laboratories for studying astrophysical phenomena such as the theoretical work on TEOs \citep{2017MNRAS.472.1538F, 2021FrASS...8...67G, 2023A&A...671A..22K}, pulsation phases and mode of TEOs \citep{2014MNRAS.440.3036O, 2020ApJ...888...95G, 2024ApJ...974..278L, 2024MNRAS.530..586L}, resonance locking \citep{2017MNRAS.472L..25F, 2018MNRAS.473.5165H, 2020ApJ...903..122C}, hybrid pulsation systems \citep{2013MNRAS.434..925H, 2019ApJ...885...46G,2023AJ....166...42W,2025PASJ...77..118L}, apsidal motion \citep{2016MNRAS.463.1199H, 2021ApJ...922...37O}, massive HBSs and exceptional objects \citep{2022A&A...659A..47K, 2024A&A...686A.199K, 2023NatAs...7.1218M, 2024A&A...685A.145K}, stellar evolution \citep{2014AA...564A..36B, 2021MNRAS.506.4083J, 2025arXiv250317133M}, and evolution of eccentric orbits \citep{2012ApJ...753...86T, 2016ApJ...829...34S, 2023ApJS..266...28L}.

It was only with the release of photometric data from long-term baseline and high-precision surveys that HBSs were discovered in large numbers. After the first Kepler HBS KOI-54 (KIC 8112039) was reported by \citet{2011ApJS..197....4W}, \citet{2016AJ....151...68K} reported the first catalog of Kepler HBSs with 173 systems. \citet{2022ApJS..259...16W} published a catalog of 991 HBSs discovered in the Optical Gravitational Lensing Experiment (OGLE; \cite{2015AcA....65....1U}) database. 

The Transiting Exoplanet Survey Satellite (TESS; \cite{2015JATIS...1a4003R}), launched in 2018, is a space-based, all-sky survey mission designed to detect exoplanets. It monitors the sky with a field of view spanning 24 $\times$ 96 degrees, observing each sector for approximately 27 days. Recent years have witnessed a growing number of HBS discoveries from TESS data. \citet{2021A&A...647A..12K} found 20 HBSs from the TESS data. \citet{2021A&A...652A.120I} reported 40 HBS candidates in their TESS eclipsing binary catalog. \citet{2022arXiv221210776B} reported a catalog of TESS variable stars, including 96 HBS candidates. \citet{2024ApJ...974..278L, 2024MNRAS.534..281L} discovered 28 TESS HBSs by visual search of light curves. \citet{2025ApJS..276...17S} identified 180 TESS HBSs. Recently, \citet{2025OJAp....8E..97C} reported 112 HBSs using GAIA and TESS data. To date, over 400 HBSs have been discovered using TESS data.

Thanks to the continued observations by the TESS survey telescope, 42 new HBSs were discovered in this work. Sect.~\ref{sect:data} describes the data reduction and analytic procedure for these objects. Sect.~\ref{sect:teos} represents the detection of the TEOs and identification of their pulsation phases and modes. Sect.~\ref{sect:pulsation} shows an HBS that exhibits gamma Doradus ($\gamma$ Dor) pulsation. Sect.~\ref{sect:discussion} analyzes the $e-P$ relation and the Hertzsprung-Russell diagram. Sect.~\ref{sect:sc} provides a summary and conclusions.

\section{Data and Modeling} \label{sect:data}
\subsection{Selection of Heartbeat Stars}
The search for HBSs using TESS data is our long-term research objective, and significant progress has been made \citep{2024ApJ...974..278L, 2024MNRAS.534..281L}. Next, we searched approximately 308,000 TESS objects in the Positions and Proper Motions (PPM) \citep{1993BICDS..42...11R, 1994A&AS..105..301R} catalog. We downloaded all the TESS photometric data for each target using the lightkurve package \citep{2018ascl.soft12013L}. Through visual inspection, we selected samples with smaller data dispersion and more prominent ``heartbeat'' signals from TESS-SPOC (the TESS Science Processing Operations Center) and QLP (the MIT Quick-Look Pipeline) data for analysis. After a visual inspection, nearly 100 HBS candidates were found.

We then excluded the high blending sectors, and removed outliers outside of 3$\sigma$ (the standard deviation of the total flux) from the average flux. We also used the Locally Weighted Scatter-plot Smoothing (LOWESS) approach \citep{cleveland1979robust} to detrend the data for samples with large deviations using OriginPro 2025b software. 

Additionally, we performed a rough data fitting to generate phase-folded light curves using the approach described in Sect. \ref{sec:modeling}. We then excluded some objects with one of the following characteristics: (a) The light curves do not fit properly. (b) The heartbeat feature becomes no longer significant in phase-folded light curves. (c) A reliable orbital period cannot be determined. Note that more accurate orbital periods require data from more sectors. For example, TIC 118196277 shows two heartbeat signals in Sector 13, and its orbital period is determined by combining light curve data from Sectors 13, 39, and 66.

Finally, we obtained these 42 HBSs shown in Table \ref{tab:hbparms}.

\subsection{The K95$^+$ model}\label{sec:modeling}
We employ the modified \citet{1995ApJ...449..294K} model (K95$^+$ model) to fit the non-eclipsing light curves. The model is characterized by seven fundamental parameters: the orbital period ($P$), eccentricity ($e$), inclination ($i$), argument of periastron ($\omega$), periastron passage epoch ($T_{0p}$), amplitude scaling factor ($S$), and fractional flux offset ($C$) \citep{2023ApJS..266...28L}. As presented in Eq. (\ref{equation:one}), this revised model incorporates a sign reversal for the $\omega$ term (from negative to positive) compared to the original formulation in Eq. (44) of \citet{1995ApJ...449..294K}, following the modification by \citet{2022ApJ...928..135W}:
\begin{equation}\label{equation:one}
	\frac{\delta F}{F}(t) = S\cdot\frac{1-3\sin^2i\sin^2(\varphi(t)+\omega)}{(R(t)/a)^3}+C,
\end{equation}
where $\delta F/F$ stands for the relative change in flux over time, $\varphi(t)$ is the true anomaly as a function of time, $R(t)$ is the distance between the components of the system as a function of time, and $a$ is the semi-major axis of the eccentric orbit. For the detailed methodology of conversion between these parameters, refer to Section 3.1 of \citet{2023ApJS..266...28L}. The model employs an approximation of the equilibrium tidal deformation with a sum of all dominant modes, with spherical harmonic degree $l=2$ and azimuthal order $m = 0, \pm2$ \citep{2021A&A...647A..12K}.

\begin{figure}
	\centering
	\includegraphics[width=0.75\textwidth]{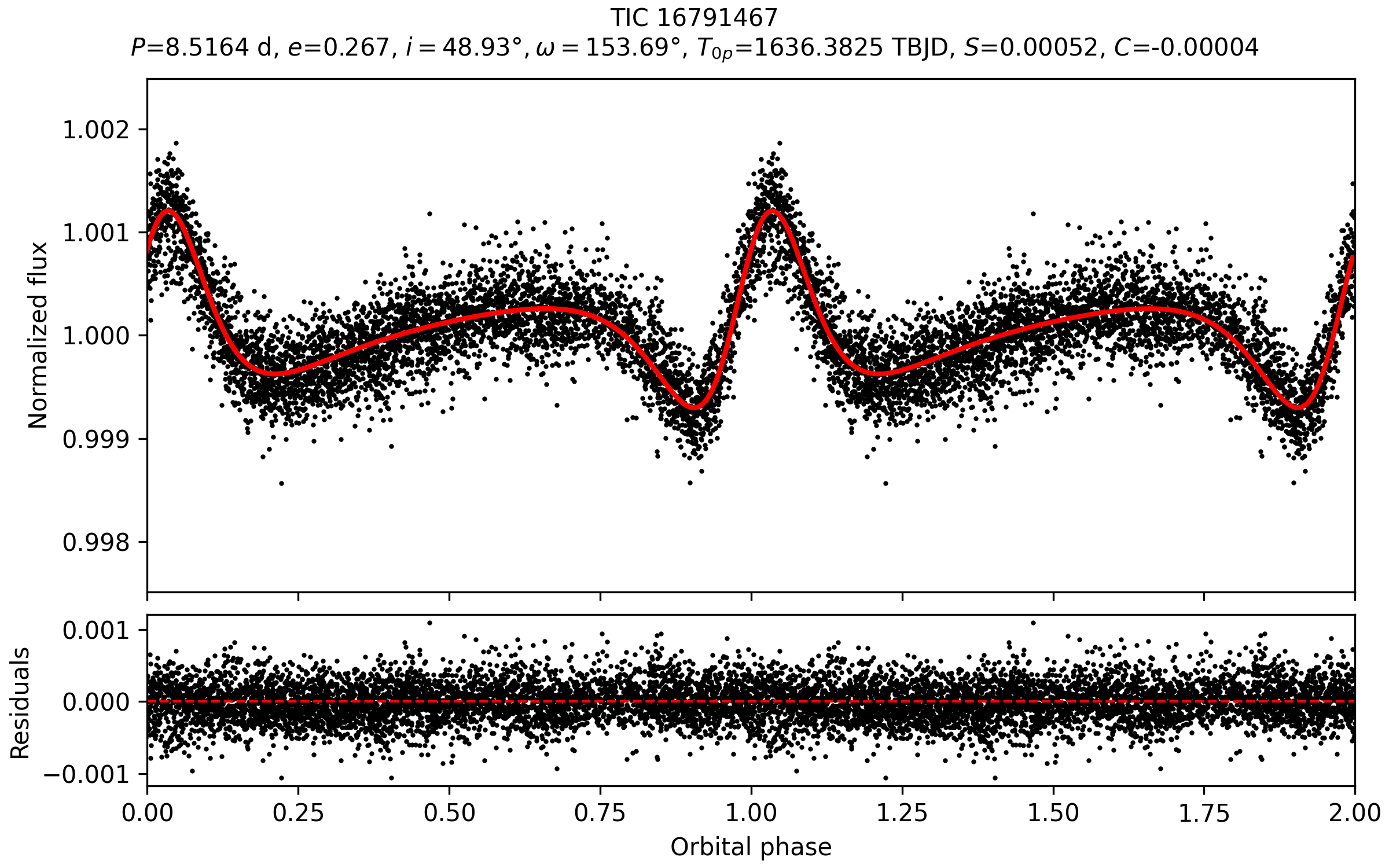}
	\caption{Fit results of TIC 16791467. The K95$^+$ model (solid red line) fitted to the phase-folded light curve (black dots) in the top panel. The lower panel shows the residuals of the fit. The dashed red line indicates a zero point. The unit of $T_{0p}$ is TBJD=BJD$-$2,457,000. Note: The plot shows phases 0$-$2 to make the fitting results clear, and phases 1$-$2 is an exact copy of phases 0$-$1.
		\label{fig:fitrst}}
\end{figure}

\begin{figure}
	\centering
	\includegraphics[width=0.75\textwidth]{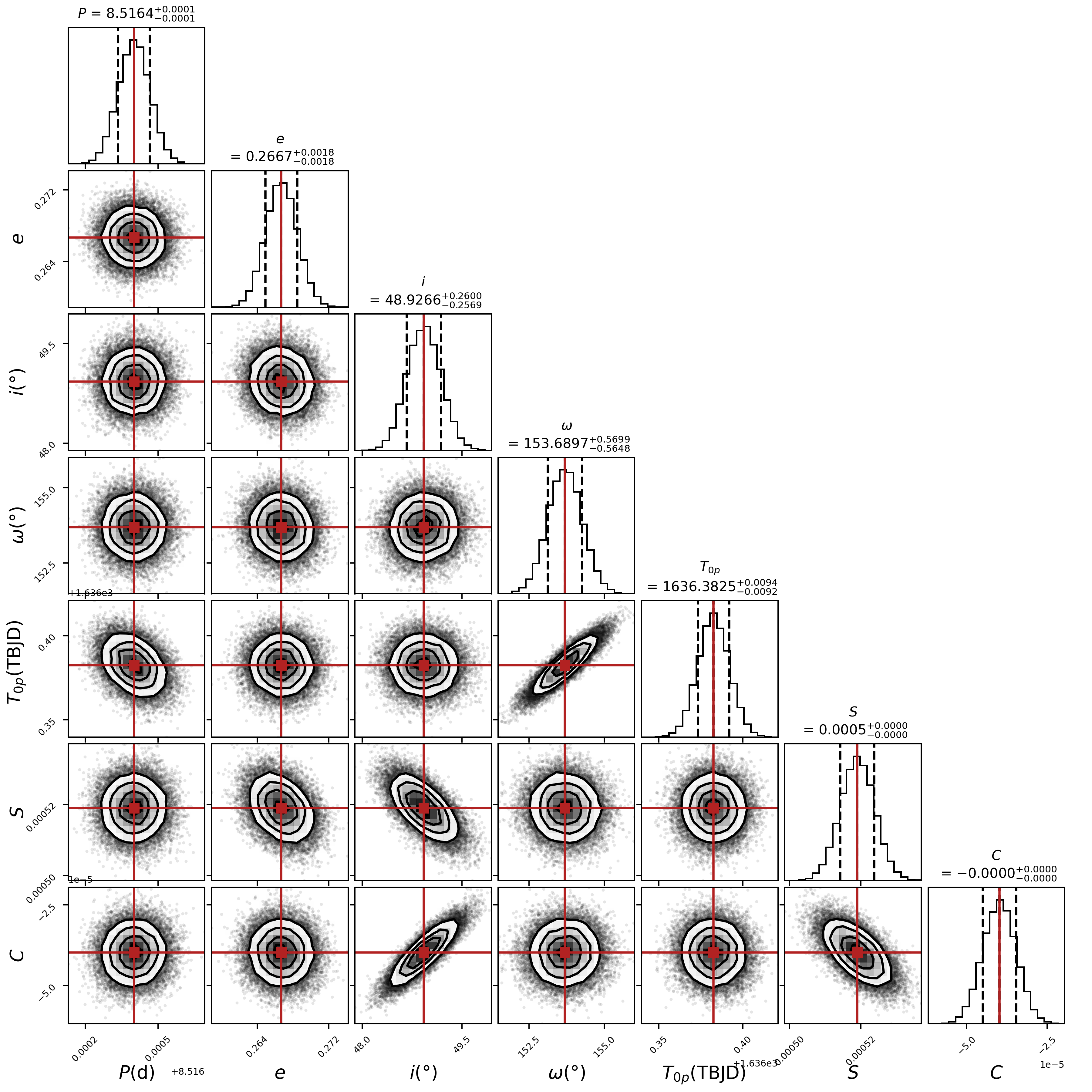}	
	\caption{Corner plot of the MCMC fit procedure for TIC 16791467. Red vertical lines indicate the median values of the presented histograms for each parameter. Black vertical dashed lines show 1 $\sigma$ uncertainties.
		\label{fig:corner}}
\end{figure}

We perform light curve fitting using the Markov Chain Monte Carlo (MCMC) technique implemented through the emcee v3.1.6 Python package \citep{2013PASP..125..306F}, adopting the methodology described in Section 3.2 of \citet{2023ApJS..266...28L}. The MCMC analysis involves two primary iterative rounds. First, we set the priors of the parameters as a uniform distribution of the physically reasonable range, and use 50 walkers and 5000 steps for a single chain to obtain the appropriate parameters for each system. Second, we set the priors of the parameters as a normal distribution centered on the value from the first round, with the standard deviation being one-hundredth of this value (for $T_{0p}$, we use $P$/100). We use 100 walkers and 10,000 steps in a single chain to estimate formal errors of the parameters. The average autocorrelation times of these systems are approximately 80 consecutive steps. Subsequently, each chain has been `thinned' by 40 steps (half of the autocorrelation time) before constructing the posterior distribution and the final estimation of the uncertainties. The parameters and their uncertainties of the HBSs are presented in Table \ref{tab:hbparms}. Figures \ref{fig:fitrst} and \ref{fig:corner} show results of the fit and the corner plot of the MCMC procedure for TIC 16791467, respectively. Plots for all samples are available as supplementary online material on China-VO at doi:\href{https://nadc.china-vo.org/res/r101666/}{10.12149/101666}.

\startlongtable
\begin{deluxetable*}{lccccccccc}
	\tabletypesize{\scriptsize}
	\tablecaption{Parameters of the K95$^+$ model fitted to the light curves of the 42 TESS HBSs.
		\label{tab:hbparms}}
	\tablehead{
		\colhead{TESS ID} & \colhead{$P$(d)} & \colhead{$e$} & \colhead{$i$($^\circ$)} & \colhead{$\omega$($^\circ$)}  & \colhead{$T_{0p}$(TBJD)} & \colhead{$S$($\times$10$^{-4}$)} & \colhead{$C$($\times$10$^{-4}$)} & \colhead{D.S.} & \colhead{sector(s)}
	}
	\startdata
	16791467 & 8.516401(66) & 0.2667(18) & 48.93(26) & 153.69(57) & 1636.3825(94) & 5.190(48) & -0.397(52) & (1) & 12,39 \\ 
	21649340 & 8.436560(48) & 0.3934(18) & 53.20(27) & 179.80(53) & 1441.7633(63) & 3.534(31) & 0.017(42) & (1) & 5,6,32,33 \\ 
	45494579 & 2.1129014(20) & 0.2454(13) & 40.28(12) & 127.79(40) & 1545.1439(19) & 7.017(49) & -2.835(43) & (1) & 9,35,36,62,63 \\ 
	54320598$^a$ & 3.8736206(10) & 0.3481(37) & 74.32(25) & 139.8(14) & 3103.027906(50) & $-$ & $-$ & (2) & 66 \\ 
	60502086 & 1.9635607(45) & 0.1297(29) & 17.13(12) & 16.65(43) & 2229.0892(14) & 42.61(95) & -35.17(87) & (1) & 34,35,61 \\ 
	91915057 & 4.0334841(33) & 0.46447(99) & 71.61(48) & 32.30(30) & 1575.0501(15) & 4.576(34) & 1.575(54) & (1) & 10,37,63,64 \\ 
	92828790 & 2.20572753(56) & 0.27318(41) & 37.623(30) & 151.32(11) & 1518.89814(47) & 21.113(39) & -9.234(33) & (1) & 8,9,35,36,61,62 \\ 
	99149998 & 8.03941(82) & 0.2599(11) & 51.09(19) & 25.10(40) & 2195.4275(68) & 16.186(97) & -2.00(11) & (1) & 7,33,34 \\ 
	118196277 & 21.725155(61) & 0.6622(11) & 61.79(42) & 113.33(60) & 1660.2273(60) & 1.100(18) & 0.145(39) & (1) & 13,39,66 \\ 
	120580114 & 10.295372(25) & 0.5625(24) & 56.48(47) & 113.24(81) & 1357.1315(64) & 1.173(28) & -0.019(36) & (1) & 2,9 \\ 
	132161007 & 4.3075569(52) & 0.3498(14) & 34.625(81) & 145.77(35) & 1545.3713(21) & 15.925(97) & -7.688(75) & (1) & 9,10,36,62,63 \\ 
	133664118 & 6.809638(12) & 0.5136(11) & 28.688(77) & 136.09(28) & 2965.4699(17) & 16.68(11) & -12.779(81) & (2) & 61,62,88,89 \\ 
	139392663 & 9.455731(42) & 0.3738(13) & 38.011(85) & 150.00(37) & 2231.8627(47) & 15.089(94) & -5.057(76) & (1) & 34,61 \\ 
	139687746 & 3.36310(20) & 0.2591(23) & 22.022(74) & 33.04(41) & 2964.4900(16) & 28.82(30) & -20.47(23) & (1) & 61,62 \\ 
	154909534 & 7.520325(27) & 0.1551(19) & 21.157(81) & 8.27(37) & 1493.1634(47) & 40.85(47) & -30.63(40) & (1) & 7,8,34,61 \\ 
	156846634 & 22.636761(100) & 0.53451(69) & 53.56(11) & 130.18(24) & 2179.5177(44) & 3.185(17) & -0.210(25) & (1) & 33,34,61 \\ 
	179741841 & 6.5114423(76) & 0.5648(10) & 52.60(16) & 96.94(35) & 1600.9274(19) & 4.864(49) & -0.716(63) & (1) & 11,38,65 \\ 
	184413651 & 4.26081881(74) & 0.50868(23) & 38.534(17) & 143.458(83) & 1518.06576(36) & 16.136(23) & -6.660(20) & (1) & 8,34,35,61,62 \\ 
	192482706 & 7.314155(12) & 0.17600(79) & 28.686(49) & 170.45(20) & 1518.9037(29) & 25.42(11) & -14.307(88) & (1) & 8,9,35,62 \\ 
	218650951 & 5.2760713(90) & 0.10712(81) & 33.11(10) & 14.04(32) & 1629.7977(43) & 36.85(22) & -19.82(21) & (1) & 12,39,66 \\ 
	260711426 & 2.9786039(59) & 0.2067(16) & 39.59(17) & 97.49(47) & 2036.0495(33) & 16.88(15) & -6.70(14) & (1) & 27,28,67,68 \\ 
	290715298 & 13.989200(46) & 0.48766(99) & 48.88(12) & 127.57(39) & 1419.9710(57) & 1.4119(94) & -0.275(12) & (1) & 4$-$7,27,31$-$34,37,64,67 \\ 
	298275012 & 11.67785(62) & 0.47879(71) & 35.303(53) & 165.97(28) & 3067.8843(30) & 3.768(14) & -1.307(11) & (1) & 65,66 \\ 
	299848020 & 11.4647(12) & 0.5303(11) & 40.47(11) & 5.14(45) & 3070.9911(36) & 10.913(64) & -2.615(88) & (1) & 65 \\ 
	302212509 & 5.0104369(51) & 0.19506(63) & 53.35(18) & 145.20(27) & 1354.8180(31) & 8.592(42) & -0.338(44) & (1) & 2,9$-$12,32,36$-$38,62$-$65 \\ 
	326691593 & 9.213305(37) & 0.3549(23) & 39.64(17) & 19.97(69) & 1658.7622(88) & 7.468(76) & -2.196(67) & (1) & 13,66 \\ 
	327328354 & 13.2056(42) & 0.21536(85) & 53.67(20) & 45.08(33) & 3066.653(12) & 14.142(87) & 0.185(85) & (2) & 65 \\ 
	336716262 & 2.29357(17) & 0.2090(11) & 33.617(93) & 114.98(29) & 3068.7672(17) & 43.94(25) & -22.01(23) & (1) & 65 \\ 
	342781296 & 9.7741(11) & 0.3445(13) & 45.35(14) & 15.21(41) & 3072.5857(63) & 11.468(68) & -1.736(75) & (1) & 65,66 \\ 
	344963507 & 3.9974559(96) & 0.1491(17) & 29.02(13) & 160.52(47) & 1629.9394(42) & 28.11(27) & -16.64(24) & (1) & 12,39,66 \\ 
	360261802 & 5.8445446(48) & 0.31520(55) & 64.58(18) & 32.97(16) & 1600.3591(16) & 11.543(42) & 2.715(52) & (1) & 11,12,38,64,65 \\ 
	362507280 & 2.908017(80) & 0.31847(72) & 28.938(45) & 111.55(12) & 3013.88745(99) & 46.61(14) & -27.62(12) & (1) & 9$-$11,36,37,63,64 \\ 
	362877322 & 8.826261(20) & 0.37455(69) & 90.00(75) & 116.07(22) & 2990.2973(27) & 9.479(30) & 2.845(46) & (2) & 62,63,89,90 \\ 
	363212626 & 7.0673(24) & 0.1327(17) & 30.31(16) & 176.51(60) & 3095.146(11) & 36.51(41) & -21.00(38) & (2) & 66 \\ 
	366958343 & 4.46609(50) & 0.5475(22) & 49.13(28) & 5.48(70) & 3095.6745(27) & 6.613(81) & -1.60(14) & (1) & 66 \\ 
	377943589 & 4.0071834(93) & 0.1913(11) & 42.08(13) & 154.45(36) & 2282.9340(33) & 16.910(95) & -4.408(92) & (1) & 36,37,63,64 \\ 
	401640445 & 7.46976(98) & 0.31668(95) & 73.98(64) & 80.58(30) & 3073.6969(37) & 9.937(88) & 3.509(80) & (1) & 65 \\ 
	445324847 & 5.2531921(34) & 0.3407(22) & 16.23(18) & 177.81(27) & 1493.48274(79) & 29.13(33) & -20.96(20) & (1) & 7$-$9,33$-$35,61,62 \\ 
	451764642 & 9.1982(12) & 0.33856(79) & 35.796(47) & 169.73(22) & 3041.2196(29) & 22.493(75) & -8.275(55) & (1) & 64 \\ 
	456588338 & 8.0262(32) & 0.4365(24) & 89.3(21) & 171.13(65) & 3091.1394(90) & 2.408(30) & 1.479(55) & (1) & 66 \\ 
	464590976 & 12.265687(23) & 0.5669(12) & 44.40(13) & 13.76(47) & 1658.7591(40) & 3.412(25) & -0.544(41) & (1) & 13,27,66,67 \\ 
	651236341 & 4.11031(25) & 0.12813(45) & 36.579(59) & 134.64(18) & 2116.2958(21) & 11.161(34) & -4.486(32) & (2) & 30 \\
	\enddata
	\tablecomments{The first column is the TESS ID; columns 2$-$8 are the seven parameters in the K95$^+$ model; column 9 shows the data source: (1) TESS-SPOC data; (2) MIT QLP data; column 10 represents the data sector(s) used for modeling. The unit of $T_{0p}$ is TBJD=BJD$-$2,457,000. \\
	$^a$ indicates that the parameters are derived using PHOEBE.\\
	(This table is available in its entirety in machine-readable form.)\\}
\end{deluxetable*}

\subsection{The PHOEBE model}\label{sec:phoebe}
In our sample, TIC 54320598 displays a single eclipse feature, rendering it incompatible with the K95$^+$ model. Consequently, we employ the PHysics Of Eclipsing BinariEs (PHOEBE) code \citep{2005ApJ...628..426P, 2016ApJS..227...29P, 2018maeb.book.....P, 2018ApJS..237...26H, 2020ApJS..247...63J, 2020ApJS..250...34C} for light curve modeling, adhering to the methodology outlined by \citet{2025PASJ...77..118L}. We first fit the eclipsing parts of the light curves to a parabola to obtain the time of superior conjunction($T_{0sc}$) and $P$. We then fix the primary effective temperature $T_{\rm eff1}$ to 9231 K, which is obtained from the {\tt\string TEFF} field in the primary header of the TESS FITS file. We also fix all bolometric albedos and gravity darkening coefficients at 1.0 according to the effective temperatures. Consequently, we will simultaneously fit seven parameters, $e$, $i$, $\omega$, $q$, $tratio$, $r_1$, and $r_2$. The MCMC run also comprises two rounds: the first round establishes appropriate parameter values within their physically reasonable ranges; the second round determines the corresponding parameter uncertainties. We use 20 walkers and 10,000 steps to obtain the parameters and uncertainties shown in Table \ref{tab:phoebe}. The average autocorrelation time is approximately 200 consecutive steps, and each chain has been `thinned' by 100 steps estimate the uncertainties.

In the absence of sufficient high-precision spectra, we could only derive these relative parameters. Nevertheless, these parameters are adequate for subsequent TEOs analysis.

\begin{deluxetable*}{lr}
	\tablewidth{0pt}
	\tablecaption{Binary Model Parameters of TIC 54320598. \label{tab:phoebe}}
	\tablehead{
		\colhead{Parameter} & \colhead{Value}
	}
	\startdata
	Orbital Period, $P$ (days) & 3.8736206(10) \\
	Time of Superior Conjunction, $T_{0sc}$ (TBJD)& 3102.768143(50) \\
	Time of Periastron passage, $T_{0p}$ (TBJD) & 3103.027906(50) $^b$ \\
	eccentricity, $e$ & 0.348(4) \\
	inclination, $i$ (degree) & 74.32(25) \\
	Argument of periastron, $\omega$ (degree) & 139.8(14) \\
	Mass ratio (M${_2}$/M${_1}$), $q$ & 0.82(12) \\
	Temperature ratio ($T_{\rm eff2}$/$T_{\rm eff1}$), tratio & 0.991(8)\\
	Primary effective temperature, $T_{\rm eff1}$ (K) & 9231 $^a$ \\ 
	Secondary effective temperature, $T_{\rm eff2}$ (K) & 9155(60)  \\
	Primary relative radius, $r_1$ (sma) & 0.1478(44) \\
	Secondary relative radius, $r_2$ (sma) & 0.1045(53) \\
	Primary bolometric albedo, $A_1$ & 1.0 $^a$\\
	Secondary bolometric albedo, $A_2$ & 1.0 $^a$\\
	Primary gravity darkening coefficient, $\beta_1$ & 1.0 $^a$\\
	Secondary gravity darkening coefficient, $\beta_2$ & 1.0 $^a$\\
	\enddata
	\tablecomments{The units of $T_{0sc}$ and $T_{0p}$ are TBJD=BJD$-$2,457,000. \\
		$^a$ fixed. \\
		$^b$ obtained from the PHOEBE Bundle after setting the $P$, $T_{0sc}$, $e$, and $\omega$ values.}
\end{deluxetable*}

\section{TEOs and phases $\&$ mode identification}\label{sect:teos}
We then detect the harmonic TEOs in these HBSs following the analytic procedure of \citet{2024ApJ...962...44L}. To extract frequencies precisely and avoid window function artifacts, we performed a Fourier analysis on the TESS data using continuous sectors (see Table 3, Column 9) and applied the FNPEAKS\footnote{\url{http://helas.astro.uni.wroc.pl/deliverables.php?active=fnpeaks}} code to compute the spectrum. The mean noise level of each frequency, N, is derived as the mean amplitude in the frequency range $\pm$1 d$^{-1}$. Only frequencies with a signal-to-noise ratio (S/N) greater than 4.0 are used for TEO analysis. If the frequency $f$ satisfies at least one of the following equations, it is considered a harmonic TEO candidate:
\begin{equation}\label{equation:a}
	|n-f/f_{\rm orb}| < 0.01,
\end{equation}
\begin{equation}\label{equation:b}
	|n-f/f_{\rm orb}| < 3\sigma_{f/f_{\rm orb}}, 
\end{equation}
where $n$ is the harmonic number, $f_{\rm orb}=1/P$ is the orbital frequency, $\sigma_{f/f_{\rm orb}}=\sqrt{P^2\sigma_f^2+f^2\sigma_P^2}$ is the uncertainty of $f/f_{\rm orb}$ according to \citet{2021A&A...647A..12K} and \citet{2022ApJ...928..135W}, $\sigma_f$ and $\sigma_P$ stand for the uncertainties of $f$ and $P$, respectively. $\sigma_f$ is estimated following \citet{2008A&A...481..571K} (their Equation (4)). $P$ and $\sigma_P$ are represented in Table \ref{tab:hbparms}. Subsequently, we detect harmonic TEOs in ten HBSs.

\begin{figure}
	\digitalasset
	\centering
	\includegraphics[width=0.5\columnwidth]{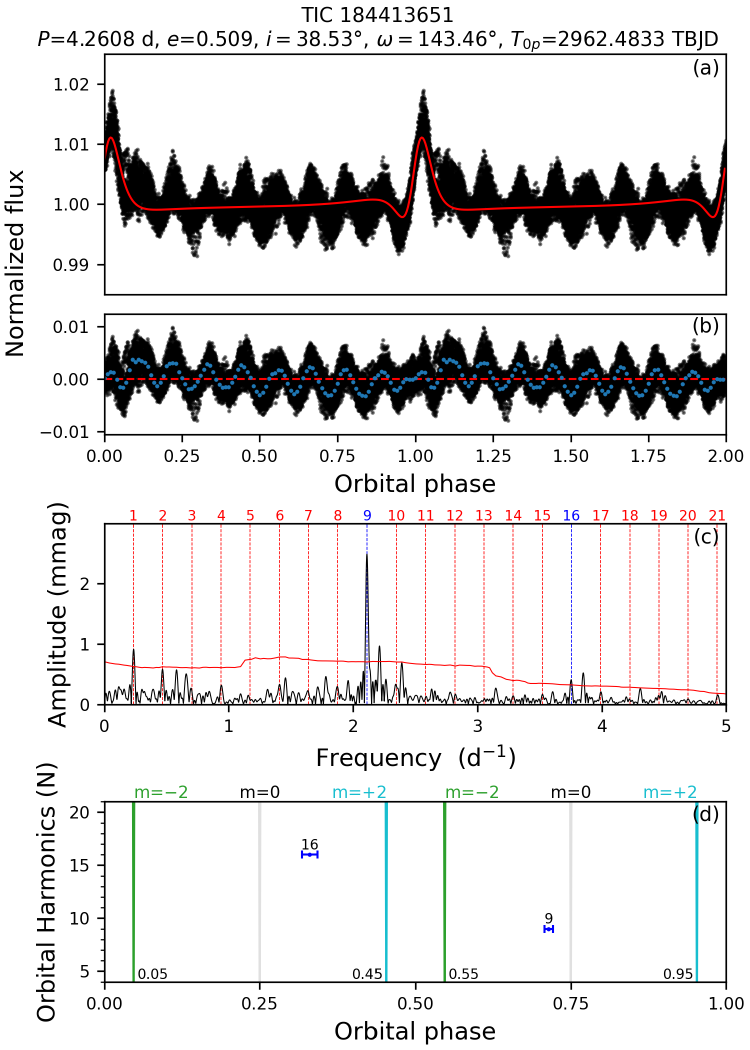}
	\caption{The analytic procedure for TIC 184413651. Panel (a): The K95$^+$ model (solid red line) fitted to the phase-folded light curve (black dots). Panel (b): The residuals of the fit in panel (a), and the blue dots are medians in 0.01 phase bins. Panel (c): The Fourier spectrum of the residuals from panel (b). The red and blue vertical dashed lines represent the orbital harmonics n; the blue lines indicate that they are harmonic TEOs. The solid red line shows the level of S/N = 4.0. Panel (d): The pulsation phases of the TEOs. The gray, light blue, and green strips indicate the $m=0,+2$, and $-2$ modes, respectively. The phases of the $m=+2,-2$ modes are shown next to the strips. The width of the strips results from the uncertainties of $T_{0p}$ and $\omega$. The blue circle represents a TEO with its harmonic number $n$; the error bar corresponds to the uncertainty of its phase. In this system, the $n$ = 9 harmonic is close to $m=0$ mode. The $n$ = 16 harmonic shows a large deviation ($>2\sigma$) from the adiabatic expectations. Given its lower amplitude, it may not be considered as a TEO candidate.
		\label{fig:184413651}}
\end{figure}

Applying the methodology of \citet{2024MNRAS.530..586L}, we then further identify the pulsation phases and modal characteristics of these TEOs. For the dominant $l=2$ spherical harmonic modes, the TEO pulsation phases are described by Eq.  (\ref{equation:phi}), which incorporates three fundamental assumptions: (1) co-alignment of the pulsation, spin, and orbital axes; (2) the pulsations are adiabatic and the TEOs are standing waves; and (3) absence of fine-tuning in the TEOs, meaning that the difference between the driving frequency and the star's intrinsic eigenfrequency is much larger than the mode damping rate. \citep{2014MNRAS.440.3036O, 2020ApJ...888...95G}:
\begin{equation} \label{equation:phi}
	\phi_{_{l=2,m}}=\left(\frac{1}{4}+m\phi_{_{0}}\right) {\rm mod} \frac{1}{2},
\end{equation}
where azimuthal order $m = 0, \pm2$, $\phi_{_{0}}=(0.25-\omega/360^{\circ}) {\rm mod} 1$; $\omega$ is the argument of periastron. All phases must be measured with respect to the epoch of periastron passage $T_{0p}$ and are in units of 360$^{\circ}$. In addition, $\omega$ and $T_{0p}$ are presented in Table \ref{tab:hbparms}. We also perform a standard prewhitening procedure using Period04 \citep{2005CoAst.146...53L}, which calculates the uncertainties according to \citet{1999DSSN...13...28M}, to derive the phases of the harmonic frequencies. The flux variation is formulated as a sinusoidal function $A_i {\rm sin} [2\pi(f_i \cdot t+\phi_i)]$, where $A_i$, $f_i$, and $\phi_i$ are the amplitude, frequency, and phase, respectively. Since $t$ is measured relative to $T_{0p}$, the time of each data point in the photometric light curves should subtract $T_{0p}$ before Fourier analysis. Finally, we identify the pulsation phases and mode of these TEOs as shown in Table \ref{tab:TEOs_Phase}. Figure \ref{fig:184413651} represents the analytic procedure for TIC 184413651 as an example.

The complete figure set (10 images) is available in the online journal.

\figsetstart
\figsetnum{3}
\figsettitle{The Phases and Mode Identification Procedure for the ten HBSs.}

\figsetgrpstart
\figsetgrpnum{3.1}
\figsetgrptitle{TIC 184413651}
\figsetplot{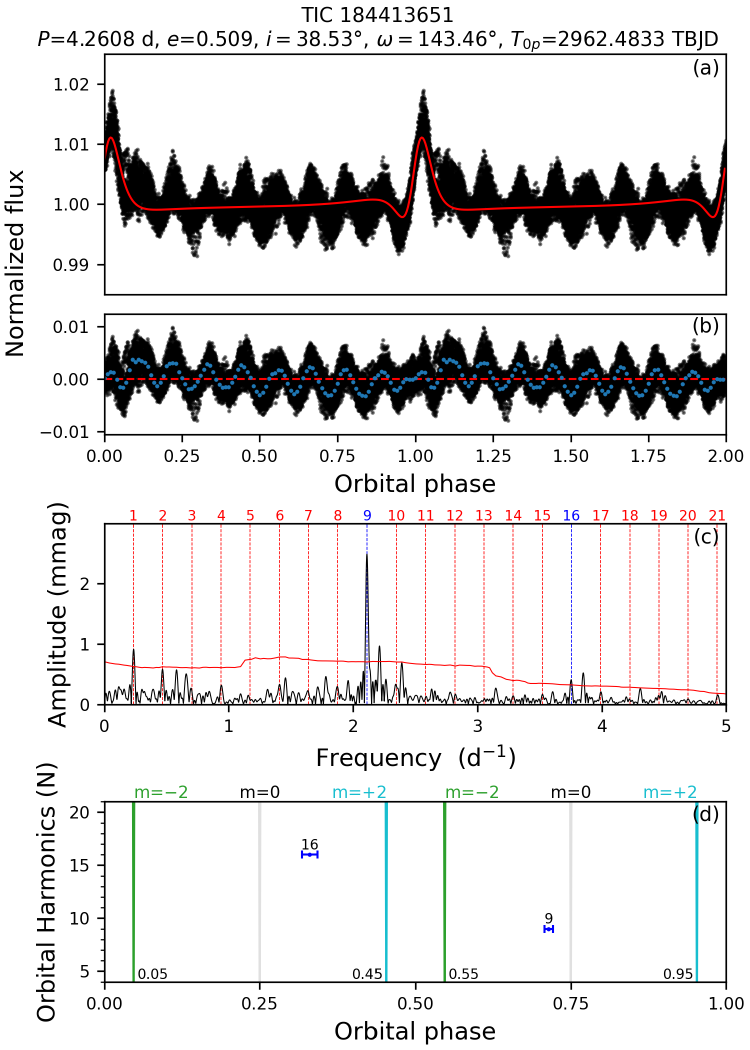}
\figsetgrpnote{The analytic procedure for TIC 184413651. Panel (a): The K95$^+$ model (solid red line) fitted to the phase-folded light curve (black dots). Panel (b): The residuals of the fit in panel (a), and the blue dots are medians in 0.01 phase bins. Panel (c): The Fourier spectrum of the residuals from panel (b). The red and blue vertical dashed lines represent the orbital harmonics n; the blue lines indicate that they are harmonic TEOs. The solid red line shows the level of S/N = 4.0. Panel (d): The pulsation phases of the TEOs. The gray, light blue, and green strips indicate the $m=0,+2$, and $-2$ modes, respectively. The phases of the $m=+2,-2$ modes are shown next to the strips. The width of the strips results from the uncertainties of $T_{0p}$ and $\omega$. The blue circle represents a TEO with its harmonic number $n$; the error bar corresponds to the uncertainty of its phase. In this system, the $n$ = 9 harmonic is close to $m=0$ mode. The $n$ = 16 harmonic shows a large deviation ($>2\sigma$) from the adiabatic expectations. Given its lower amplitude, it may not be considered as a TEO candidate.}
\figsetgrpend

\figsetend

\startlongtable
\begin{deluxetable*}{ccccccccccc}
	\tabletypesize{\scriptsize}
	\label{tab:TEOs_Phase}
	\tablecaption{TEO and phases $\&$ mode identification results of the ten TESS HBSs.}
	\tablehead{
		\colhead{TESS ID}& \colhead{$n$} & \colhead{$\Delta n$} &  \colhead{$3\sigma-|\Delta n|$} & \colhead{Frequency} & \colhead{Amplitude}  & \colhead{Phase} & \colhead{$S/N$}& \colhead{Sector(s)} & \colhead{Cadence} & \colhead{Comments} \\
		{} & {} & {} &{} & \colhead{(day$^{-1}$)} & \colhead{(mmag)} &{} &{} &{} &{s}
	} 
	\startdata
	54320598 & 10 & 0.003 & 0.050997 & 2.5828(21) & 0.505(31) & 0.073(10) & 4.75 & 66 & 200 & $m=-2$ mode\\ 
	91915057 & 12 & -0.010 & 0.007406 & 2.9729(13) & 0.0674(83) & 0.123(20) & 4.69 & 63,64 & 200 & $m=2$ mode \\  
	91915057 & 17 & 0.007 & 0.011361 & 4.2164(16) & 0.0525(83) & 0.063(25) & 4.03 & 63,64 & 200 & $m=2$ mode\\
	118196277 & 13 & 0.102 & 0.063128 & 0.60283(63) & 0.354(11) & 0.555(5) & 4.82 & 39 & 600 & $m=2$ mode\\ 
	118196277 & 24 & -0.023 & 0.143736 & 1.10453(79) & 0.286(11) & 0.757(6) & 4.75 & 39 & 600 & $m=0$ mode\\ 
	139392663 & 6 & -0.008 & 0.056278 & 0.63353(49) & 0.561(12) & 0.756(3) & 6.16 & 61 & 200 & $m=0$ mode\\ 
	156846634 & 11 & -0.001 & 0.078183 & 0.48599(32) & 0.1893(58) & 0.853(5) & 5.61 & 33,34 & 600 & (1)\\
	179741841 & 19 & -0.012 & 0.032607 & 2.9163(14) & 0.268(15) & 0.656(9) & 6.58 & 65 & 200 & $m=2$ mode\\
	184413651 & 9 & 0.003 & 0.002981 & 2.112021(77) & 2.475(17) & 0.714(1) & 14.16 & 61,62 & 200 & $m=0$ mode\\
	184413651 & 16 & -0.009 & 0.007461 & 3.75428(42) & 0.449(17) & 0.330(6) & 5.54 & 61,62 & 200 & (2)\\ 
	362507280 & 4 & 0.002 & 0.003797 & 1.37595(26) & 0.3761(95) & 0.266(4) & 9.59 & 63,64 & 200 & $m=0$ mode\\
	362507280 & 5 & 0.004 & 0.004393 & 1.72061(47) & 0.2100(95) & 0.879(7) & 6.22 & 63,64 & 200 & $m=-2$ mode\\
	377943589 & 6 & -0.001 & 0.009550 & 1.49726(29) & 0.2776(78) & 0.706(4) & 7.57 & 63,64 & 200 & $m=0$ mode\\
	464590976 & 37 & -0.037 & 0.002192 & 3.01435(64) & 0.1344(83) & 0.798(10) & 6.65 & 66,67 & 200 & $m=-2$ mode\\
	464590976 & 36 & 0.041 & 0.002887 & 2.93600(82) & 0.1051(83) & 0.896(13) & 5.15 & 66,67 & 200 & $m=-2$ mode\\
	\enddata
	\tablecomments{The first column is the TESS ID; $n$ is the harmonic number of the TEO; $\Delta n=f/f_{\rm orb}-n$, where $f$ is the detected frequency (column 5); a positive value in column 4 indicates that it satisfies Eq. (\ref{equation:b}); column 6 is the amplitude; column 7 is the phase; $S/N$ is the signal-to-noise ratio; column 9 shows the sector(s) used for the TEO analysis; column 10 shows the data sampling cadence; column 11 represents the phase identification results. Additional notes:(1) traveling wave or significantly nonadiabatic; (2) TEO-mimicking signal.}
\end{deluxetable*}

\section{HBS with $\gamma$ Dor Pulsation}\label{sect:pulsation}
\begin{figure}
	\centering
	\includegraphics[width=1.0\textwidth]{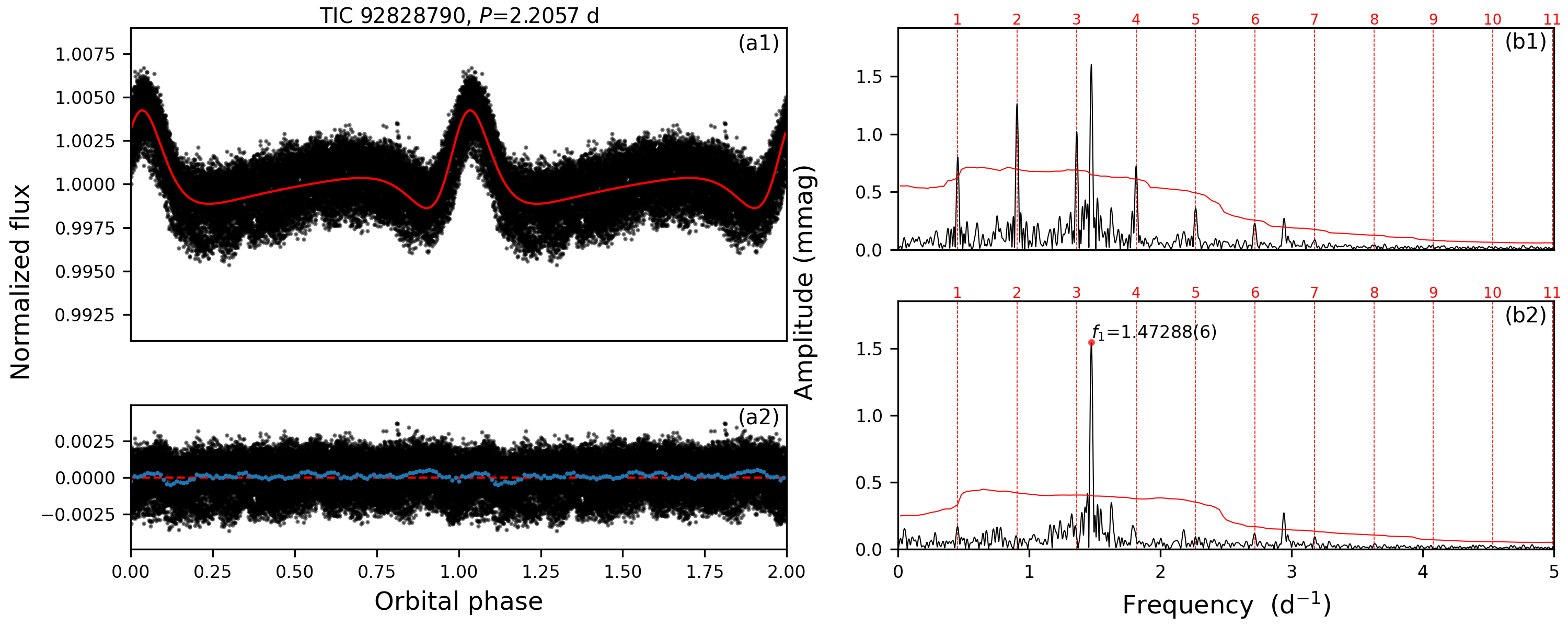}
	\caption{The analytic procedure for TIC 92828790. Panel (a1): The K95$^+$ model (solid red line) fitted to the phase-folded light curve (black dots). Panel (a2): The residuals of the fit in panel (a1). The blue dots are medians in 0.01 phase bins. Panel (b1): The frequency spectrum of the light curve shown in panel (a1). Panel (b2): The frequency spectrum of the residuals shown in panel (a2). Panels (b1) and (b2): The solid red line shows the amplitudes at $S/N$ = 4.0 as a function of frequency; the red vertical dashed lines represent the orbital harmonics $n$.
		\label{fig:11403032-ft}}
\end{figure}

During the TEO analysis of these systems, we found that while TIC 92828790 shows no TEOs, it exhibits a significant non-harmonic frequency at $f_1$ = 1.47288(6) (as shown in panel (b2) of Figure \ref{fig:11403032-ft}). We also derive the pseudo-synchronous rotation frequency of this system using the following equation \citep{1981A&A....99..126H}:
\begin{equation}\label{equation:psrot}
	f_{\rm ps\text{-}rot} = \frac{1+\frac{15}{2}e^2+\frac{45}{8}e^4 +\frac{5}{16}e^6 }{(1+3e^2+\frac{3}{8}e^4){(1-e^2)}^{3/2}} f_{\rm orb}.
\end{equation}
We then obtain $f_{\rm ps\text{-}rot}$=0.66101(65). Given the pulsation frequencies of $\gamma$ Dor are mainly around 0.3 to 3.3 d$^{-1}$, this significant frequency $f_1$ is most likely a $\gamma$ Dor pulsation. In addition, Sect.~\ref{subsec:HR} will show that this system is located precisely in the $\gamma$ Dor region of the Hertzsprung-Russell diagram. Therefore, the significant frequency $f_1$ is a $\gamma$ Dor pulsation candidate. HBSs hosting $\gamma$ Dor and/or $\delta$ Sct pulsations are exceptionally rare. To the best of our knowledge, only four HBSs with these characteristics have been reported to date \citep{2013MNRAS.434..925H, 2019ApJ...885...46G, 2023AJ....166...42W, 2025PASJ...77..118L}. TIC 92828790 will thus serve as a valuable research target in this field for further study.

\section{Discussion}\label{sect:discussion}
\subsection{The Eccentricity–Period Relation} \label{subsec:relation}
\begin{figure}
	\centering
	\includegraphics[width=0.5\hsize]{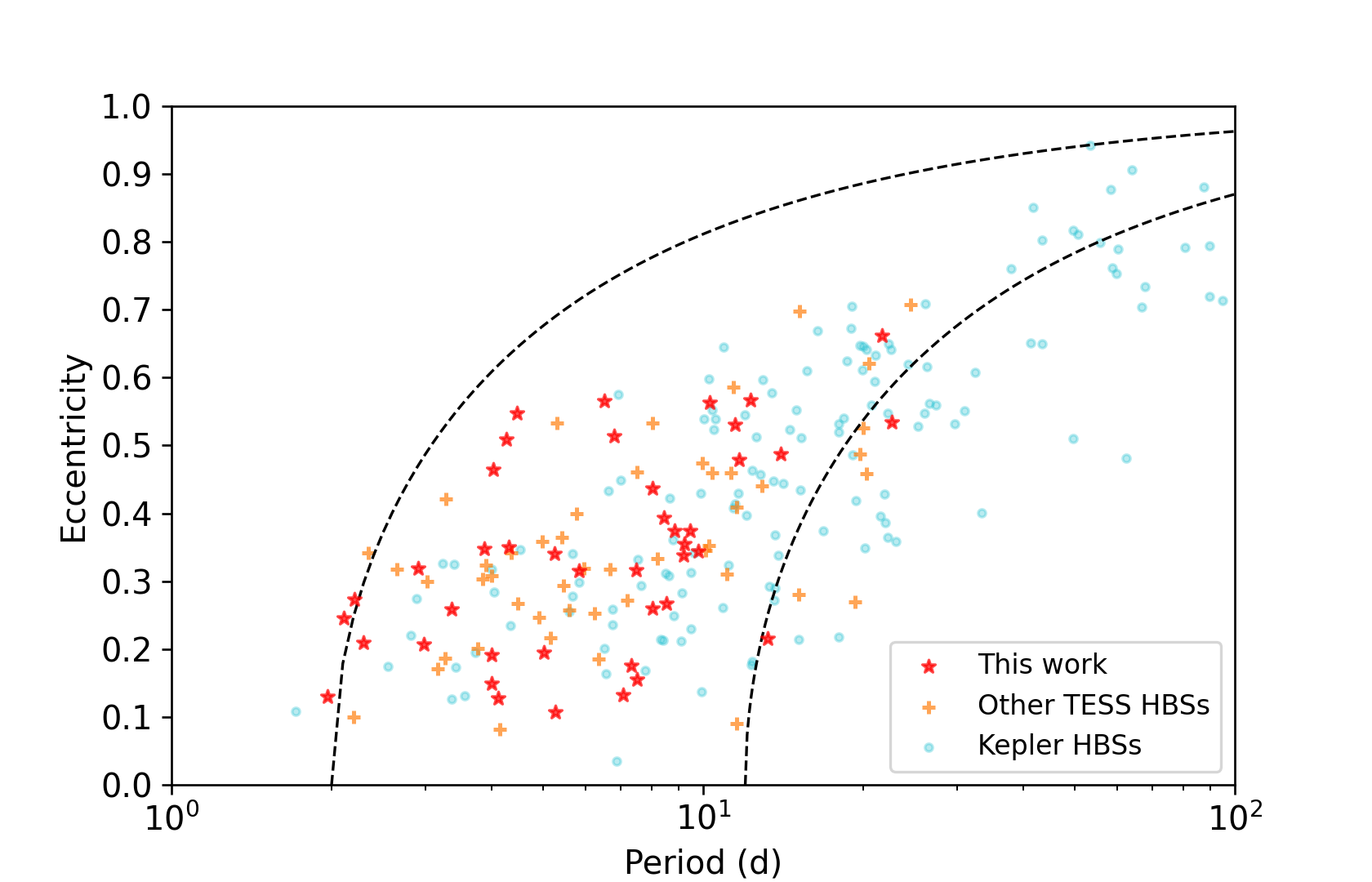}
	\caption{The eccentricity-period ($e-P$) diagram. Red stars indicate the TESS HBSs in this work. Orange pluses indicate the TESS HBSs from \citet{2021A&A...647A..12K, 2024ApJ...974..278L, 2024MNRAS.534..281L}. Blue circles represent the positions of the Kepler HBSs reported in \citet{2023ApJS..266...28L}. The two black dashed curves mark an eccentricity-period relation of $e=\sqrt{1-(P_0/P)^{2/3}}$, which is the expected functional form assuming conservation of angular momentum. The two curves use $P_0$ of 2 and 12 days.
		\label{fig:P_e}}
\end{figure}

Figure \ref{fig:P_e} shows the eccentricity-period ($e-P$) diagram. The red stars represent the 42 TESS HBSs in this work. The orange pluses indicate the TESS HBSs from \citet{2021A&A...647A..12K, 2024ApJ...974..278L, 2024MNRAS.534..281L}. The blue circles are the Kepler HBSs reported in \citet{2023ApJS..266...28L}. The two black dashed curves represent the expected function of the eccentricity-period relation assuming conservation of angular momentum \citep{2016ApJ...829...34S}: $e=\sqrt{1-(P_0/P)^{2/3}}$. The range of circularization periods ($P_0$) is 2 to 12 days, which differs from the range reported for Kepler HBSs \citep{2023ApJS..266...28L}. This difference implies selection effects in the orbital period distribution of TESS HBSs.

Furthermore, the $e-P$ distribution demonstrates a significant positive correlation for the majority of systems \citep{2023ApJS..266...28L}. This distribution provides clear evidence of ongoing orbital circularization in HBSs, as previously reported by \citet{2012ApJ...753...86T, 2016ApJ...829...34S}. The observed trend reveals that the shorter-period orbits are likely to circularize more quickly since they are undergoing stronger tidal forces; the eccentricity is smaller for shorter-period systems.

There are two systems with relatively long orbital periods: TICs 118196277 and 156846634, with periods of 21.7 and 22.6 days, respectively. However, it remains challenging to detect HBSs with longer orbital periods using TESS data because consecutive heartbeat signals must be observed across multiple continuous sectors.

\subsection{The Hertzsprung-Russell Diagram of the HBSs}
\label{subsec:HR}
To illustrate the Hertzsprung-Russell (H-R) diagram, we obtain the visual magnitude, the interstellar extinction, and the interstellar extinction of these HBSs from the Gaia Survey, which provides very high-precision astrometric data for nearly 2 billion stars \citep{2016A&A...595A...1G, 2018A&A...616A...1G, 2021A&A...649A...1G}. We query the geometric distance of these objects according to \citet{2021AJ....161..147B}. We also use the effective temperature obtained from the {\tt\string TEFF} field in the primary header of the TESS FITS file. The luminosities of these systems are calculated using the following equations:
\begin{equation}\label{equation:c}
	{\rm log}(L/L_{\odot})=0.4\cdot(4.74-M_V-BC)
\end{equation}
\begin{equation}\label{equation:d}
	M_V=m_V-5\cdot {\rm log}(D_{ist})+5-A_V,
\end{equation}
where the bolometric correction BC is estimated following \citet{2023ApJS..265...33S} and \citet{2013ApJS..208....9P}; $m_V$ is the visual magnitude, $A_V$ is the interstellar extinction, and $D_{ist}$ is the geometric distance.

\begin{figure}
	\centering
	\includegraphics[width=0.5\hsize]{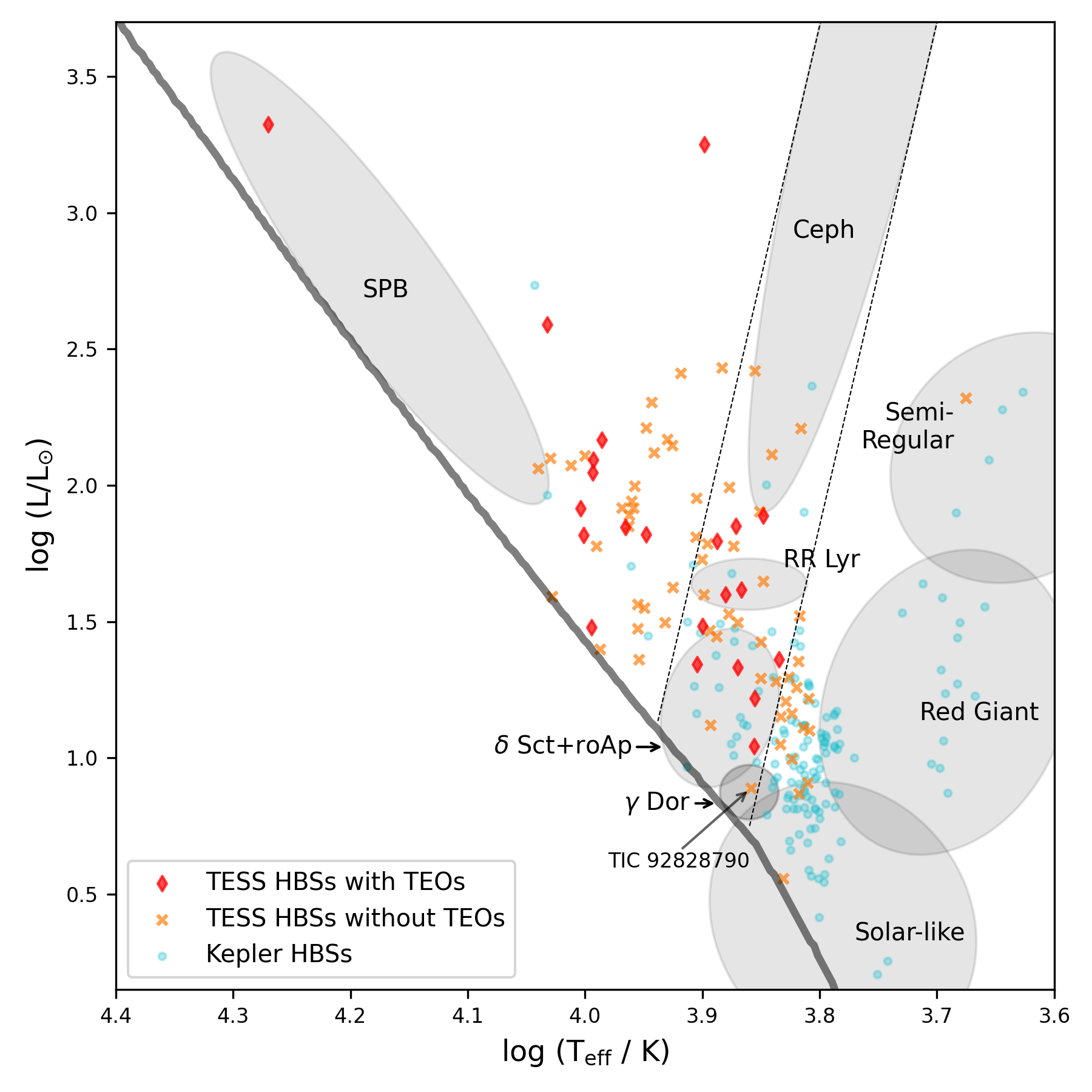}
	\caption{The H-R diagram of the HBSs. Red diamonds and orange forks represent all TESS HBSs from \citet{2021A&A...647A..12K, 2024ApJ...974..278L, 2024MNRAS.534..281L} and this work. The red diamonds represent HBSs with TEOs. The orange forks represent the HBSs without TEOs. The blue circles represent the Kepler HBSs from \citet{2023ApJS..266...28L}. The bold black solid curve represents the ZAMS. The gray dashed curves show the classical instability strip for radial pulsations. The classes of pulsating variables, including slowly pulsating B (SPB) stars, Cepheids (Ceph), RR Lyr, $\delta$ Sct+rapidly-oscillating Ap (roAp) stars, $\gamma$ Dor, semi-regular, red giant, and solar-like, are labeled next to the corresponding gray regions. All the regions are plotted according to \citet{2013PhDT.........6P, 2019ApJS..243...10P, 2021RvMP...93a5001A}.
		\label{fig:HR-all}}
\end{figure}

Figure \ref{fig:HR-all} shows the H-R diagram of all TESS HBSs in this work and from \citet{2021A&A...647A..12K, 2024ApJ...974..278L, 2024MNRAS.534..281L}, as well as the Kepler HBSs from \citet{2023ApJS..266...28L}. As can be seen, the HBSs can appear in a much wider area of the H-R diagram. Meanwhile, these TESS HBSs are located in regions with higher effective temperatures and luminosities, which may be an observational effect. Unlike Kepler, TESS is more likely to detect HBSs with higher masses and temperatures because it covers a wider survey area and can detect brighter stars.

\subsection{The reason for only one eclipse}
\begin{figure*}
	\centering
	\includegraphics[width=0.8\hsize]{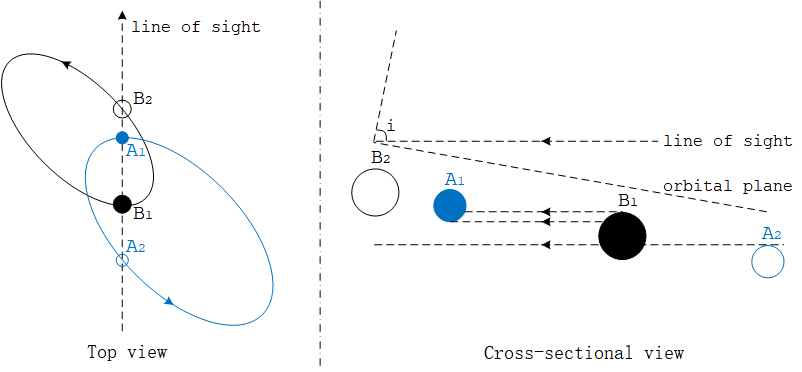}
	\caption{The eccentric orbit and the only one eclipse of the two components.}
	\label{fig:orbit}
\end{figure*}

As shown in Fig.\,\ref{fig:orbit}, the unique geometry of eccentric orbits determines that only one eclipse occurs under certain conditions. In the top view, the blue star moves to A$_1$, while the black star moves to B$_1$. At this point, the two stars align with the line of sight. As the blue star moves to A$_2$ and the black star moves to B$_2$, they realign with the line of sight again. The key difference is that the distance between the two stars differs for these two alignments. As shown in the cross-sectional view, when the inclination falls within a certain range of angles (but not excessively large), an eclipse occurs only at positions A$_1$ and B$_1$, but not at A$_2$ and B$_2$. Thus, under these inclination conditions, only one eclipse event is observed. Figure 13 of \citet{2025ApJS..276...17S} also supports this reasoning by showing the inclination distribution.

Additionally, the value of $\omega$ also determines the distance between the two components of the two line-of-sight alignments, thereby influencing the eclipse conditions. \citet{2025ApJS..276...17S} also noted this point. However, they further suggested that in some cases, the heartbeat signal might eliminate the eclipse signal, resulting in only one observable eclipse.  We argue that this scenario is unlikely because the photometric morphologies of the heartbeat and eclipse signals are distinct, making mutual elimination implausible. 

\section{Summary and conclusions}\label{sect:sc}
Based on TESS data, we have discovered 42 new HBSs. We modeled their light curves using the K95$^+$ or PHOEBE models and estimated their orbital parameters. Ten of the HBSs exhibit TEOs, and their pulsation phases and modes have been identified. The geometric effect of the dominant $l=2$, $m=0$, or $\pm2$ modes can explain most pulsation phases of the TEOs, assuming pulsations are stand waves and adiabatic. The pulsation phase of the 11th harmonic of TIC 156846634 shows large deviation from the adiabatic expectation. This suggests that it is either a traveling wave or significantly nonadiabatic. The $n$ = 16 harmonic in TIC 184413651 should not be considered a TEO candidate due to its large deviation from the adiabatic expectation. Moreover, TIC 92828790 is a candidate system for a HBS exhibiting $\gamma$ Dor pulsation.

The $e-P$ relation also shows a positive correlation \citep{2023ApJS..266...28L}, and the existence of orbital circularization \citep{2012ApJ...753...86T, 2016ApJ...829...34S}. We identify two systems with relatively long orbital periods: TICs 118196277 and 156846634, exhibiting periods of 21.7 and 22.6 days, respectively. Due to observational duration limitations of TESS Sectors, detecting HBSs with even longer orbital periods remains challenging.

In the H-R diagram, TESS HBSs exhibit higher temperatures and greater luminosities compared to Kepler HBSs, which may be attributed to observational effects. They are primarily A- and F-type stars, which are either on the main sequence or have just evolved off it. This advantage enables TESS data to observe more massive HBSs while also facilitating the detection of TEOs. 

Since \citet{2021A&A...647A..12K} reported the first TESS HBSs catalog, the sample size in the TESS HBSs catalog has been steadily increasing \citep{2022arXiv221210776B, 2024ApJ...974..278L, 2024MNRAS.534..281L, 2025ApJS..276...17S, 2025OJAp....8E..97C}. The HBSs discovered in this work may serve as valuable additions to the TESS HBS catalog.


\begin{acknowledgments}
This work is supported by the Yunnan Fundamental Research Projects (grant Nos. 202501AS070055, 202401AS070046, 202503AP140013, 202301AT070352), the International Partnership Program of Chinese Academy of Sciences (grant No. 020GJHZ2023030GC), the China Manned Space Program with grant No. CMS-CSST-2025-A16, the CAS ``Light of West China'' Program, the Basic Research Project of Yunnan Province (grant No. 202201AT070092), the Yunnan Revitalization Talent Support Program, the National Natural Science Foundation of China (No. 12573038), the China Postdoctoral Science Foundation under Grant Number 2025M773194, and the Postdoctoral Fellowship Program of CPSF under Grant Number GZC20252095. The NASA Explorer Program provides funding for the Kepler and TESS missions. The Gaia Survey is a cornerstone mission of the European Space Agency. We thank the Kepler, TESS, and Gaia teams for their support and hard work. We would like to thank the anonymous referees for many helpful comments and suggestions which improved the quality of this manuscript.
\end{acknowledgments}

\software{lightkurve \citep{2018ascl.soft12013L},
          emcee \citep{2013PASP..125..306F},
          PHOEBE \citep{2005ApJ...628..426P, 2016ApJS..227...29P, 2018maeb.book.....P, 2018ApJS..237...26H, 2020ApJS..247...63J, 2020ApJS..250...34C},
          FNPEAKS (Z. Koo{\l}aczkowski, W. Hebisch, G. Kopacki),
          PERIOD04 \citep{2005CoAst.146...53L},
          OriginPro 2025b software (OriginLab Corporation, Northampton, MA, USA).
          }




\bibliography{sample7}{}
\bibliographystyle{aasjournal}



\end{document}